\documentclass[12pt]{article}
\usepackage{latexsym,amsfonts,amsmath,amsthm,amssymb, bm}
\title{Bounds for the Distance Dependence of Correlation Functions of Entangled Photons in Waveguides}

\author{A. Khrennikov, B. Nilsson, S. Nordebo\\ International Centre for Mathematical Modelling\\ in Physics and Cognitive Sciences\\ School of Computer Science, Physics and Mathematics \\Linn\ae us University, SE-3159 V\"axj\"o, Sweden\\
I.V. Volovich\\Steklov Mathematical Institute\\ Russian Academy of Sciences \\Gubkin St. 8, 117966, GSP-1, Moscow, Russia}

\begin{document}
\maketitle

\begin{abstract}
The distance dependence of the probability of observing two photons in a waveguide is investigated. The Glauber  correlation functions of the entangled photons  in waveguides are considered and the spatial and temporal dependence of the correlation
functions is evaluated. We derive upper bounds to the distance dependence of the probability of observing two photons. These inequalities should be possible to observe in experiments.
\end{abstract}







\section{Introduction}
The transmission of light in waveguides, in particular its quantum properties, is a topic of great interest in optics. Investigations of quantum correlations and entanglement among photons have been in focus of in the foundations of quantum theory and its applications to quantum information science and metrology.

With the emergence of  quantum communication links over
long distances \cite{LSS,LHB,UTS,HVL}, there is a need for
a detailed study of the dependence of correlations
and entanglement among photons on distance.

Since direct interactions between photons in free space are
extremely weak, generation of correlated photons generally
requires nonlinear media such as the parametric down conversion.
Recently, studies of two-photon scattering from a two-level system
inside a one-dimensional  waveguide have  reported various
features of photon correlation \cite{KHT,SF,SF2,Roy}. In particular a
formal scattering theory to study multi-photon transport in a
waveguide was employed \cite{SS}. We also mention a general model
describing nonlinear effects in propagation to and from the system
in the quantum  state, based on one-dimensinal model of field-atom
interaction \cite{HOFF}, see even \cite{HOFF1}. It describes
spatiotemporal quantum coherence for the case of spontaneous
emission from a single excited atom. In \cite{KHT} this model was
applied to the two-photon input wave packets. This field of
research, spatiotemporal behavior of correlations of two photons
propagating in nonlinear media, is closely related to studies on
nonlinear response of a single atom to an input of two photons,
e.g., from a single photon source \cite{Yam}. This response can be
observed in the correlations between the two output photons. Here
it is also very important to understand spatiotemporal dependence
of correlations.

There are thus sound reasons to study how correlation functions of
entangled photons in hollow  waveguides behave in space and time.
The physical mechanism to model is dispersion that spreads pulses
in space and time causing attenuation with distance. The spreading
limits also the bit rate for a given waveguide length because of
mixing of pulses. Repeaters can be used at some length intervals,
which cause higher costs and problems with preserving the quantum
state through the repeater. Information on space and time
properties of the correlation functions is thus of engineering
interest. Such information is provided in the current paper by
modelling the effect of modal dispersion.

The study is also a preparation for more elaborate models,
including material dispersion in hollow waveguides and fibres, cf.
e.g. \cite{BB}, \cite{BB1}. To describe the situation, a brief
background is presented \cite{Agrawal2002}  on optic fibres. For a
fibre, it is possible to purify the material to the extent that
losses from scattering from the impurities can be neglected in
some wavelength bands. Examples of such bands are the 1.3 $\mu$m
and 1.55 $\mu$m bands. The dispersion can be reduced to a
negligible level for a certain wavelength within such a band, and
a low dispersion can then be reached for a narrow wavelength band.
Such a reduction of dispersion can be reached \cite{Agrawal2002}
with a dispersion compensator which is a special fibre with tuned
length, after the transmission fibre, having the opposite
dispersion to the transmission fibre. Another method
\cite{Agrawal2002} is dispersion shift, i.e., a choice of the
transversal fibre dimensions. The chromatic dispersion, defined as
the combination of modal and material dispersion, can be made to
vanish for both methods at the design wavelength.

It is worth mentioning the existence of repeaters for classical optic fibres using optical rather than previously used electrical methods \cite{Kartalopoulos2002} and fibre switches that preserve the quantum state of the photon \cite{HallAltepeterKuma2010}.

We shall study the asymptotic behaviour of the Glauber correlation
functions for the entangled states of two photons  in waveguides
and show their vanishing for large distances. We estimate the rate
of decrease of correlations and present upper bounds for the
correlation functions.

To estimate the correlation function we shall use some results on
the properties of solutions of the  (1+1)-dimensional Klein-Gordon
equation analogously used in the Haag-Ruelle scattering theory.

We prove that the  probability density $P(z_1,t_1,z_2,t_2)$ observing one photon at point $z_1$ along the waveguide at time $t_1$ and another photon at point $z_2$ at time $t_2$ satisfies
the following inequality
\begin{equation}\label{bound}
P(z_1,t_1,z_2,t_2)\leq\frac{C}{(t_0+|t_1|)(t_0+|t_2|)}
\end{equation}
for some $t_0$ and all $z_1,t_1,z_2,t_2$.

The bound (\ref{bound}) and the bounds (\ref{K-G222}), (\ref{K-G223}) (see below) should be possible to observe in experiments.

Note that the decreasing of the correlations for entangled states
in empty space was found in \cite{Vol},  see also discussion in
\cite{Vol1}, \cite{OV}. In this paper we have considered the
waveguides and found the universal bound for the correlations, see
also \cite{OLD} for preliminary considerations.

\section{The photon probability density}
Let $E_j(\bm{r},t)$ be the $j$-th component $(j=1,2,3)$ of the
electric field operator at the space time point ${\bf r},t$. The
operator can be written as the sum of the positive and negative
frequency parts:
$E_j({\bm r},t)=E_j^{(+)}({\bm r},t)+ E_j^{(-)}({\bf r},t).
$
The probabilities of photo detection are given by Glauber`s
formulas, \cite{MW}. In particular, the probability that a
state $\psi$ of the radiation field will lead to the detection at
time $t$ of a photon with the polarization along the direction $j$
by a detector atom placed at point ${\bm r}$ is
proportional to the first-order
correlation function $P_{\psi}({\bm r},t,j)=\langle\psi|E_j^{(-)}({\bm r},t)
E_j^{(+)}({\bm r},t)|\psi\rangle.
$
The joint probability of observing one photo ionization with
polarization $j_{1} $ at point ${\bm r}_{1}$ at time between
$t_{1}$ and $t_{1} +dt_{1}$ and another one with polarization
$j_{2}$ at point ${\bm r}_{2}$ between $t_{2}$ and $t_{2} +dt_{2}$
with $t_{1}\leq t_{2}$ is proportional to
$P_{\psi}({\bm r}_{1},t_{1},j_{1};{\bm r}_{2},t_{2},j_{2})dt_{1}dt_{2}$,
where the second-order correlation function is
\begin{equation}
\label{i3}P_{\psi}({\bm r}_{1},t_{1},j_{1};{\bm r}_{2},t_{2},j_{2})
=\langle\psi|E_{j_{1}}^{(-)}({\bm r}_{1},t_{1})
E_{j_{2}}^{(-)}({\bm r}_{2}%
,t_{2})
 E_{j_{2}}^{(+)}({\bm r}_{2},t_{2})
E_{j_{1}}^{(+)} ({\bm r}_{1}%
,t_{1})|\psi\rangle
\end{equation}

The waveguide is a hollow conducting cylindrical tube
$\Gamma\subset \mathbb{R}^{3}$ along the $z$ axis with boundary
surface $S$ and with a cross section $\Omega$ with the bounding
curve $\partial\Omega$ in the $xy$-plane. It will be assumed that
the walls have infinite conductivity. Appropriate boundary
conditions are posed:
$E_{t}|_{S}=0,~~H_{n}|_{S}=0,
$
where $E_{t}$ is the component of electric field ${\bm E}$ tangential to the
boundary of the waveguide and $H_{n}$ is the component of magnetic field ${\bm H}$
normal to the boundary.

It is well known that in
the interior of the waveguide the solutions of the Maxwell equations
without sources can be divided into two sets of solutions, the so
called $TM$ modes with $H_{z}=0$ and the $TE$ modes with $E_{z}=0$
\cite{LL,Col}. A general solution is a linear combination of the
$TM$ and $TE$ modes.

The general solution of the Maxwell equations in the waveguide can
be written as follows \cite{Kri}. Let $\varphi_{n\nu}(z,t)$ be any
function satisfying the Klein-Gordon equation
\begin{equation}\label{GS1}
(\frac{\partial^{2}}{\partial t^{2}}-\frac{\partial^{2}}
{\partial z^{2}%
}+m_{n\nu}^{2})\varphi_{n\nu}(z,t)=0.%
\end{equation}
Here $n=1,2,...$ and $\nu=TM$ or $TE$. We define for $\nu=TM$
\begin{equation}\label{GS2}
{\bm E}_{n\nu}({\bm r},t)={\bm e}_{n\nu}(x,y)m_{n\nu}^{-1}%
\frac{\partial}{\partial z}\varphi_{n\nu}(z,t)\end{equation}$$
+{\bm n}_{z}m_{n\nu}v_{n}%
(x,y)\varphi_{n\nu}(z,t),$$%
\[
{\bm H}_{n\nu}({\bm r},t)=-{\bm h}_{n\nu}(x,y)m_{n\nu}^{-1}%
\frac{\partial}{\partial t}\varphi_{n\nu}(z,t)
\]
where $v_n$ is the solution of the eigenvalue problem

\begin{equation}
\label{s4}(\frac{\partial^{2}}{\partial x^{2}}+\frac{\partial^{2}}{\partial
y^{2}} +m_{n}^{2})v_{n}(x,y)=0,~~~~(x,y)\in\Omega,
\end{equation}
\[
v_{n}|_{\partial\Omega}=0
\]
with the properties
\begin{equation}
\label{s5}\int_{\Omega}v_{n}(x,y)v_{n^{\prime}}(x,y)dxdy=\delta_{nn^{\prime}},
\end{equation}
\begin{equation}
\label{s6}\sum_{n}v_{n}(x,y)v_{n}(x^{\prime},y^{\prime})=\delta(x-x^{\prime})
\delta(y-y^{\prime}),
\end{equation}
where $m_{n}^{2}>0$ are the eigenvalues.

It is defined for $\nu=TM$
\begin{equation}
\bm{e}_{n\nu}(x,y)=\nabla_{T}v_{n}(x,y), \label{g5}%
\end{equation}%
\[
\bm{h}_{n\nu}(x,y)=\bm{n}_{z}\times\nabla_{T}v_{n}(x,y)=\bm{n}_{z}%
\times\bm{e}_{n\nu}(x,y).
\]
${\bm E}_{n\nu}({\bm r},t)$ and ${\bm H}_{n\nu}({\bm r},t)$ are defined in an analogous manner \cite{OLD} for $\nu=TE$.

The general solution of the Maxwell equations in the waveguide
 can now be written in the form
\begin{equation}
\label{g8}{\bm E}({\bm r},t) =\sum_{n\nu}{\bm E}_{n\nu}%
({\bm r},t),
{\bm H}({\bm r},t) =\sqrt{\frac{\epsilon_0}{\mu_0}}
\sum_{n\nu}{\bm H}_{n\nu}({\bm r},t).
\end{equation}

We write the solution of the Klein-Gordon Eq. (\ref{GS1}) in the form
\begin{equation}
\label{s6a}\varphi_{n} (z,t)=\int\frac{d k}{2\sqrt{2\pi\omega_{n}(k)}}%
(a^{+}_{n}(k) e^{i\omega_{n}(k) t-ikz}\end{equation} $$+a_{n}(k)
e^{-i\omega_{n}(k) t+ikz}),$$
where $\omega_{n}(k)=\sqrt{k^{2}+m_{n}^{2}},
$
and quantize it by taking $a_{n}(k), a_{n}^{+}(k)$ as the annihilation and creation operators.

Now the quantum electromagnetic field in the waveguide is reduced
to a set of massive (1+1)-dimensional  Klein-Gordon fields. Let us
consider one of the modes. We define a one particle state
\begin{equation}\label{onep}
|\psi_1\rangle=\int g(k)a_{k}^{\dag}|0\rangle,
\end{equation}
where $|0\rangle$ is the Fock vacuum.
The probability density to detect the photon at the  point
$z$ along the waveguide at time $t$ is proportional to
\begin{equation}\label{pro1}
P(z,t)=\langle \psi_1|\varphi^{(-)}(z,t)
\varphi^{(+)}(z,t)
|\psi_1\rangle,
\end{equation}
where
\begin{equation}\label{fie}
\varphi^{(+)} (z,t)=
\frac{1}{(2\pi)^{1/2}}\int_{\mathbb{R}}
\frac{dk}{\sqrt{2\omega_k}}
a_ke^{-i\omega_kt+ikz},\varphi^{(-)}(z,t)=h.c.,
\end{equation}
and $\omega_k=\sqrt{k^2+m^2}$,\,$m>0$.

The expression (\ref{pro1}) can be written as
\begin{equation}\label{pros12}
P(z,t)=|A(z,t)|^2,
\end{equation}
where
\begin{equation}\label{pros13}
A(z,t)=\langle 0|\varphi^{(+)}(z,t)
|\psi_1\rangle=
\int
dk
g(k)\frac{e^{-i\omega_{k}t+ikz}}
{2\sqrt{2\pi\omega_{k}}}.
\end{equation}
The function $A(z,t)$ is a solution of the Klein-Gordon equation. Let us consider
the question of how the solution behaves in the limit of large $t$ in a frame of reference
moving with a constant velocity $v\leq V<1$.  In this frame, corresponding to the time
$x(1-v)/v$ after the wave front at $t=x$, the field is given by
\begin{equation}\label{pros14}
A(vt,t)=
\int dk
g(k)\frac{e^{-it(\omega_{k}-kv)}}
{2\sqrt{2\pi\omega_{k}}}.
\end{equation}
For sufficiently smooth function $g(k)$
one can use the stationary phase method to get
\begin{equation}\label{pros15}
A(vt,t)=
g(k_0)\frac{e^{-it(\omega_{k_0}-k_0v)-i\pi/4}}
{2\sqrt{2\pi\omega_{k_0}}}\sqrt{\frac{2\pi}{t\omega^{''}(k_0)}}
+O(\frac{1}{t}).
\end{equation}
Here $k_0$ is the solution of the equation $\omega^{'}(k)=v$, i.e. $k_0=mv/ \sqrt{1-v^2}$ and one has $\omega^{''}(k_0)>0$.
For the probability $P(z,t)$ we obtain
\begin{equation}\label{pros17}
P(vt,t)=
\frac{1}{t}\frac{|g(k_0)|^2}
{4\omega_{k_0}\omega^{''}(k_0)}
+O(\frac{1}{t^{3/2}}).
\end{equation}
More elaborated results on asymptotic expansions of the solution to the Klein-Gordon equation are given by H\"ormander \cite{Hor}.

\section{The entangled photon correlation function with distance dependence}
Now we define a two particle entangled state (biphoton)
\begin{equation}\label{ent}
|\psi\rangle=\int f(k_1,k_2)a_{k_1}^{\dag}a_{k_2}^{\dag}|0\rangle,
\end{equation}
where $|0\rangle$ is the Fock vacuum and $f(k_1,k_2)$
is the two-photon wave function which is a symmetric function, $f(k_1,k_2)=f(k_2,k_1)$ because we deal with bosons.

The probability to detect one particle at the  point
$z_1$ along the waveguide at time $t_1$ and another particle at the space point $z_2$ at time $t_2$ is proportional to
\begin{equation}\label{pro}
P(z_1,t_1,z_2,t_2)=\langle \psi|\varphi^{(-)}(z_1,t_1)\varphi^{(-)}(z_2,t_2)
\varphi^{(+)}(z_2,t_2)
\varphi^{(+)}(z_1,t_1)
|\psi\rangle.
\end{equation}

The expression for $P(z_1,t_1,z_2,t_2)$ (\ref{pro})
can be written as
\begin{equation}\label{pros1}
P(z_1,t_1,z_2,t_2)=|A(z_1,t_1,z_2,t_2)|^2,
\end{equation}
where
\begin{equation}\label{pros2}
A(z_1,t_1,z_2,t_2)=\langle 0|\varphi^{(+)}(z_1,t_1)
\varphi^{(+)}(z_2,t_2)
|\psi\rangle=
\int
dk_1dk_2\end{equation}
$$
\{\frac{e^{-i\omega_{k_2}t_2+ik_2z_2}}
{2\sqrt{2\pi\omega_{k_2}}}
\frac{e^{-i\omega_{k_1}t_1+ik_1z_1}}{2\sqrt{2\pi\omega_{k_1}}}
f(k_1,k_2)+(k_1\leftrightarrow k_2)\}.
$$
Note that the function $A(z_1,t_1,z_2,t_2)$ satisfies the Klein-Gordon
equation with respect to $z_1,t_1$ and $z_2,t_2$.

Let us suppose that one of the photons is observed in a frame of
reference moving with velocity $v_1$  and another photon is
observed in a frame of reference moving with velocity $v_2$. By
using the stationary phase method we obtain for large $t_1$ and
$t_2$:
\begin{equation}\label{pros21}
A(z_1,t_1,z_2,t_2)=
\sqrt{\frac{2\pi}{t_1\omega^{''}(k_{10})}}
\sqrt{\frac{2\pi}{t_2\omega^{''}(k_{20})}}
f(k_{10},k_{20})\frac{1}{2\sqrt{2\pi\omega_{k_{10}}}}
\frac{1}{2\sqrt{2\pi\omega_{k_{20}}}}e^{-i\pi/2}
\end{equation}
$$
\{  e^{-it_1(\omega_{k_{10}}-ik_{10}v_1)}
e^{-it_2(\omega_{k_{02}}-ik_{20}v_2)}
+(k_1\leftrightarrow k_2)\}.
$$
Here $k_{10}=mv_1/ \sqrt{1-v_1^2}$, $k_{20}=mv_2/ \sqrt{1-v_2^2}$, $z_1=v_1t_1$ and $z_2=v_2t_2$.

Therefore
\begin{equation}\label{pros12}
P(z_1,t_1,z_2,t_2)=
\frac{|f(k_{10},k_{20})|^2}
{16t_1t_2\omega^{''}(k_{10})\omega^{''}(k_{20})
\omega_{k_{10}}\omega_{k_{20}}}.
\end{equation}
$$
|\{  e^{-it_1(\omega_{k_{10}}-ik_{10}v_1)}
e^{-it_2(\omega_{k_{02}}-ik_{20}v_2)}
+(k_1\leftrightarrow k_2)\}|^2
$$
It is interesting to see the difference between the entangled wave function $f(k_{10},k_{20})$ and the separable one by looking to it with an explicitly indicated dependence on the spacetime coordinates:
\begin{equation}\label{pros123}
f(k_{10},k_{20})=f(m\frac{z_1}{t_1}/\sqrt{1-
\frac{z_1^2}{t_1^2}},m\frac{z_2}{t_2}/\sqrt{1-
\frac{z_2^2}{t_2^2}})
\end{equation}

 Let the wave function of two photons $f(k_1,k_2)$ in a waveguide be a smooth fast decreasing function. Then  the probability of observing two photons in the waveguide should satisfy the following bounds. For any $n_1,n_2=0,1,2,3,...$ there exist constants $C_{n_1n_2}$ such that for
\begin{equation}\label{K-G222}
|z_1|\geq |t_1|,\,|z_2|\geq |t_2|,
\end{equation}
one has
\begin{equation}\label{K-G223}
P(z_1,t_1,z_2,t_2)
\leq\frac{C_{n_1n_2}}{(1+|z_1|)^{n_1}(1+|z_1|)^{n_2}}.
\end{equation}
Furthermore there exists a constant $C$ such that
\begin{equation}\label{K-G321}
P(z_1,t_1,z_2,t_2)\leq\frac{C}{|t_1||t_2|},
\end{equation}
for all $z_1,t_1,z_2,t_2$. An asymptotic estimate for $C$, valid for large $t_1$ or $t_2$, is provided by (\ref{pros12}).

An explicit expression for the wave function of the biphotons in a special case
is given in \cite{YLS}:
\begin{equation}\label{expl}
f(k_1,k_2)=\frac{i}{k_0^2}f_P(k_1+k_2)\sqrt{6k_1k_2(k_1+k_2)},
\end{equation}
 where $f_P(k_1+k_2)$ is a Gaussian function describing the pumping photons. By using this form of the wave
 function we obtain the asymptotic formula for the probability in this special case.

 To conclude,  the main result of this paper is the bounds
 to the probability density  (\ref{bound}) and  (\ref{K-G222}), (\ref{K-G223}) which, in principle,  should be possible to test in experiments.

 {\bf Acknowledgements}. One of the authors (I. Volovich) would like to thank the International Centre for Mathematical Modelling in Physics and Cognitive Sciences for the support during his visit to Linn\ae us University.


\end{document}